\documentclass[prd,twocolumn,nofootinbib,superscriptaddress]{revtex4-1}
\input epsf
\usepackage{graphics}
\usepackage{amsmath,amssymb}
\usepackage{color}
\usepackage{dcolumn}
\usepackage{hyphenat}
\usepackage{multirow}

\usepackage{graphicx}

\def\be{\begin{equation}}
\def\ee{\end{equation}}
\def\ba{\begin{eqnarray}}
\def\ea{\end{eqnarray}}

\newcommand{\beqa}{\begin{eqnarray}}
\newcommand{\eeqa}{\end{eqnarray}}

\newcommand{\beq}{\begin{equation}}
\newcommand{\eeq}{\end{equation}}

\newlength{\tskip}
\newlength{\colwidth}

\begin{document}

\title{Matching LIGO/Virgo merger rates with primordial black holes formed during the QCD epoch }
\title{Evidence for primordial black hole dark matter from LIGO/Virgo merger rates}

\author{Karsten Jedamzik} 
\affiliation{Laboratoire de Univers et Particules de Montpellier, UMR5299-CNRS, Universite de Montpellier, 34095 Montpellier, France}


\begin{abstract}
The LIGO/Virgo collaboration has by now observed or constrained the 
gravitational merger rates of different classes of compact objects.
We consider the possibility that the bulk of these mergers are 
primordial black hole (PBH) mergers of PBHs formed during the QCD epoch
making up the entirety of the dark matter.
Having shown in a companion paper that mergers due to the initial binary
population formed in the early Universe are likely 
negligible, we compute current
merger rates in PBH clusters in which the typical PBH resides.
We consider two scenarios: (i) the PBH mass function dictated by the
QCD equation of state and (ii) the PBH mass function dictated by the 
existence of a peak in the inflationary perturbation spectrum. In the first
scenario, which is essentially parameter free, we reproduce very well
the merger rates for heavy BHs, the merger rate of mass-asymmetric BHs
such as GW190814, a recently discovered merger of a 23$M_{\odot}$ black 
hole with a $2.6M_{\odot}$ object, 
and can naturally explain why LIGO/Virgo has not yet observed mergers
of two light PBHs from the dominant $\sim 1\,M_{\odot}$ PBH population.
In the second scenario, which has some parameter freedom, 
we match well the observed rate of heavy PBHs, but
can currently not explain the rate for mass-asymmetric events. In either 
case it is tantalizing that in both scenarios PBH merger rates made with
a minimum of assumptions match most LIGO/Virgo observed rates very well.      
\end{abstract}

\maketitle

With the technological advancement of the LIGO and Virgo gravitational
wave detectors a new field of observational astrophysics/cosmology has
emerged. LIGO/Virgo by now routinely observe the gravitational wave emission
of mergers of black holes (BH) or neutron star (NS) binaries out to very
large cosmic distances. Each observational run seems to provide a new surprise,
LIGO O1 - the detection of a very massive binary of two $\sim 30M_{\odot}$
BHs~\cite{Abbott:2016blz}, O2 - 
the detection of many such massive binaries with often close to
equal masses of the binary components and very low spin
~\cite{LIGOScientific:2018mvr,LIGOScientific:2018jsj}, and O3 - the
detection of a binary merger, with one of the binaries seemingly being too
massive for a NS and too light for being a 
BH~\cite{Abbott:2020khf}. These observations have 
confronted the theoretical community of stellar population models with new
challenges. It seems not impossible that $\sim 30M_{\odot}$ may be produced
as a result of star formation (and evolution) if it happens in a low-metallicity environment~\cite{Belczynski:2009xy,Belczynski:2010tb}, 
or $\sim 2.6M_{\odot}$ BHs are made by the 
merger of two lighter NSs as in the case of GW170817 (see ~\cite{Broadhurst:2020cvm,Safarzadeh:2020ntc} for alternative explanations). 
 
In this letter, we follow an alternative path, investigating if
the observed objects could be primordial black holes (PBH) formed during the
QCD epoch (for general reviews on PBHs 
cf.~\cite{Khlopov:2008qy,Carr:2020gox}). 
It has been known for some while that equation of state (EOS,
hereafter) effects
during the QCD epoch lead to an enhanced PBH formation on a particular
scale~\cite{Jedamzik:1996mr,Jedamzik:1998hc,Jedamzik:1999am}, provided the underlying inflationary density fluctuations are
approximately scale-invariant. Recent investigations of the resulting
mass spectrum~\cite{Byrnes:2018clq,Carr:2019kxo,Sobrinho:2020cco}, under the assumption of Gaussianity for the fluctuations,
and taking into account the accurate zero chemical potential
EOS~\cite{Borsanyi:2016ksw,Bhattacharya:2014ara}, reveal a well-defined peak on a scale somewhere around
$0.5 - 2M_{\odot}$ containing most of the PBHs and a shoulder between
$\sim 8-50 M_{\odot}$ 
containing a smaller mass fraction $f_M\approx 10^{-2}$ of PBHs. Here the
uncertainties are mostly due to the semi-analytic nature of the estimates, 
and may well be addressed by complete general-relativistic numerical 
simulations of the PBH formation process~\cite{Niemeyer:1997mt,Niemeyer:1999ak,Musco:2004ak}. 
An alternative scenario of PBH formation during the QCD epoch may be 
imagined if inflation, accidentally has left a peak in the power spectrum
on the required scale, for example due to a temporary flattening of the inflaton potential~\cite{Ivanov:1994pa,Bullock:1996at} (see ~\cite{Dolgov:1992pu} for another alternative).
Many inflationary scenarios leading to PBH formation
have been investigated, too numerous to cite here.
In what follows we will investigate these two alternative explanations
for the LIGO/Virgo observations. 
Our complete attention is on the prediction of binary
merger rates.
Throughout this paper we will assume that PBHs contribute the entirety of the cosmic dark matter.

It had been believed for some time now that $\sim 1-30M_{\odot}$ PBHs can not contribute
the entirety of the dark matter, rather only a very small fraction, as it
was argued that very eccentric and hard binaries forming shortly before
cosmological matter-radiation equality would lead to current BH-BH merger rates 
incompatible with those inferred by 
LIGO/Virgo~\cite{Sasaki:2016jop,
Ali-Haimoud:2017rtz,Ballesteros:2018swv,Bringmann:2018mxj,Raidal:2018bbj,
Vaskonen:2019jpv,DeLuca:2020qqa}.
In a recent paper, appearing two days before the announcement of GW190814, we have shown that the conclusions about
the viability of PBH dark matter are drastically modified when considering
three body interactions between the two binary members and a third 
by-passing PBH in early PBH clusters~\cite{Jedamzik:2020ypm}. 
In fact, we have found for $M = 1M_{\odot}$ that the formerly predicted approximate merger rate
${\cal M}_{M_{\odot}}\approx 1.25\times 10^6{\rm Gpc^{-3}yr^{-1}}$ is reduced to ${\cal M}\approx 40\,{\rm Gpc^{-3}yr^{-1}}$ by three-body
interactions. The rate is likely even lower if more frequent encounters
are taken into account.
This rate should be compared to the current upper limit
${\cal M}_{M_{\odot}}\approx 5.2\times 10^3{\rm Gpc^{-3}yr^{-1}}$ from LIGO/Virgo. It is important to note that the
prediction of these rates is subject to one important uncertainty, i.e.
the fraction of binaries which never entered clusters, $f_{\rm free}$. 
We will return to this point below.

Building on prior results~\cite{Afshordi:2003zb,Chisholm:2005vm,
Raidal:2018bbj,Inman:2019wvr} 
(cf. also to~\cite{Trashorras:2020mwn}), we have argued 
in ~\cite{Jedamzik:2020ypm} that PBHs form 
clusters at very high redshifts which evaporate later due to the continuous
loss of PBHs in the high energy tail of the Maxwell-Boltzman 
energy distribution. We have found that clusters of $N_{cl}\approx 1300$
with approximate densities $n_{cl}\approx 3.6/{\rm pc^3}$
having formed at redshifts $z\approx 85$ are evaporating at the current 
epoch. We will use these values as reference values for the typical
environment in which PBHs are at the present epoch. However, all given 
merger rates will include the full dependence on clusters
properties, given by the
PBH number in a cluster $N_{cl}$ and the PBH masses.
\vskip 0.15in
{\it PBH formation during the QCD epoch with scale-invariant inflationary perturbations:}
\vskip 0.15in

We stress here that when referring to scale-invariant fluctuations we do
not imply that these fluctuations have the same amplitude as on the much larger cosmic microwave background scales, but rather are 
approximately scale-invariant over the 2-3 efolds during inflation
which produce perturbations on the QCD scales considered here.
We will approximate the mass function of such PBHs
by a simple bi-modal distribution with a peak mass 
scale $M_s$ containing the majority of PBHs, i.e. $f_M(M_s)\approx 1$
and a larger mass scale $M_b\gg M_s$ containing a much smaller mass fraction
$f_M(M_b)\ll 1$. Given the simplicity of our estimate any more detailed
treatment is currently not warranted.
We now proceed to the calculation of merger rates.
We will assume that, as has been
explicitly shown in the case of $M_s$ - $M_s$ PBH binaries, with $M_s = M_{\odot}$, 
that three-body PBH interactions will significantly modify the initial 
eccentricity $e$ and semi-major axis $a$ distributions $P(a,e)$ for
$M_s$-$M_b$ and $M_b$-$M_b$ binaries, such that their rates are
well below the current inferred rates for such mergers. These assumptions
seem very reasonable also since for typical $f_M(M_b)\approx 10^{-2}$
there should be much fewer $M_b$-$M_b$ binaries than when 
$f_M(M_b)\approx 1$. Similar arguments apply to the frequency of
$M_s$-$M_b$ binaries. With the initial binary population disrupted,
and the subsequent formation of very hard and eccentric binaries
improbable, the current merger rate would we dominated by direct capture
of two single PBHs.

\begin{figure}[htbp]
\centering
\includegraphics[width=0.48\textwidth]{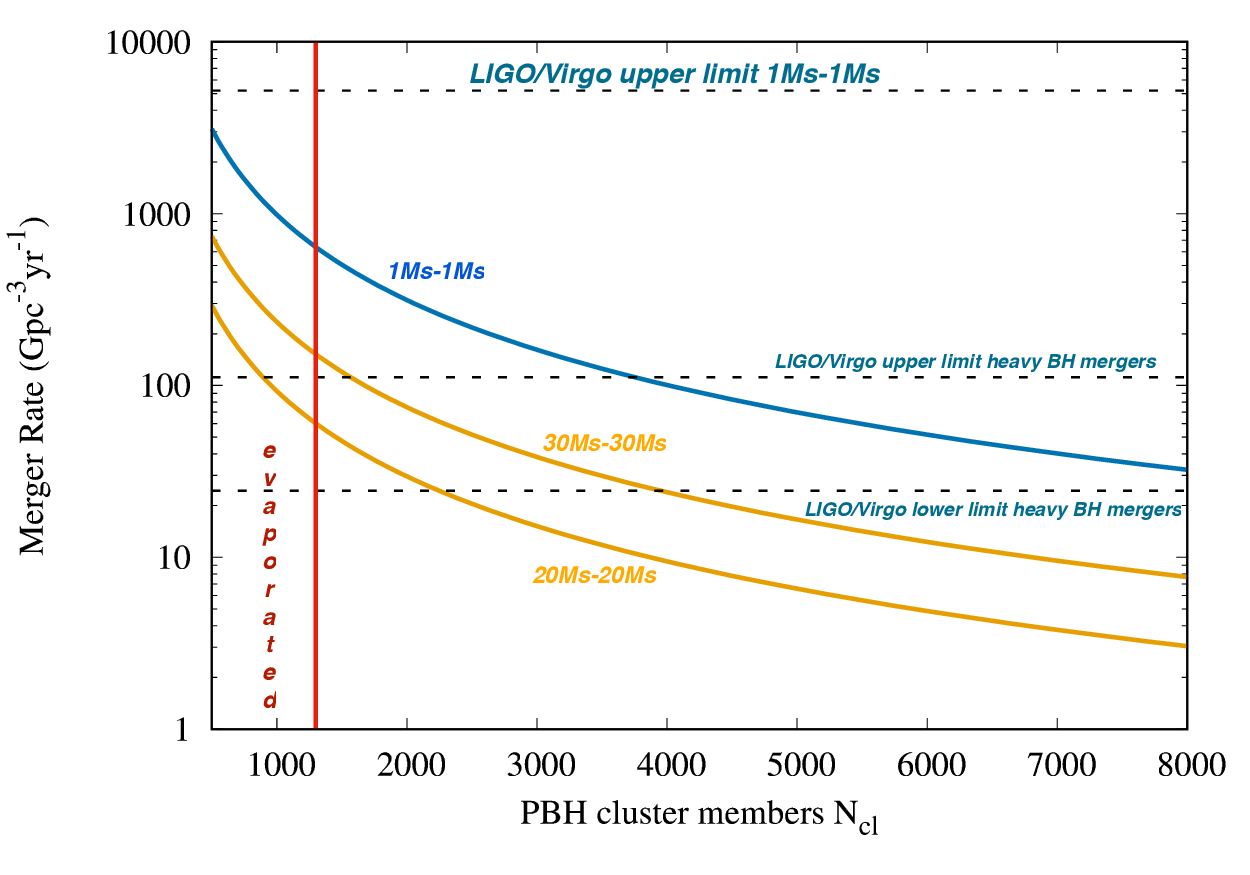}
\caption{\label{fig:merge_large} 
The predicted current merger rates
of two equal mass PBHs as a function of the number of PBH cluster 
members $N_{cl}$ they reside in today. Results for unpeaked inflationary
perturbations are shown. 
Small PBH clusters with $N_{cl}$
left to the vertical red line have evaporated by today~\cite{BT08} 
and don't contribute. The upper horizontal dashed line shows the $90\%$
confidence upper limit of the LIGO/Virgo
rate on coalescence of two $1\, M_{\odot}$ BHs and should be compared
to the dark blue line given by Eq.~(\ref{eq:1_1}). The lower two dashed horizontal lines show the $90\%$ confidence LIGO/Virgo range of the
detected merger rate of heavy PBHs. 
These should be
compared to the two orange lines (cf. Eq.~(\ref{eq:30_30_EOS})), showing the PBH merger rate range 
for heavy $20\, M_{\odot}\, <\, M_{pbh}\, <\, 30\, M_{\odot}$ PBHs.
It is important to note that most PBHs are expected to reside in PBH
clusters just right to the red line, above the current cluster
evaporation limit. 
}
\end{figure}

Single PBHs may coalesce by emission of gravitational radiation
during close encounters. The fraction of PBHs which merge in this 
way over a time interval $\Delta t$ is given by
\begin{equation}
f_{direct} = \frac{\Delta n_{pbh}}{n_{pbh}}
= \frac{1}{2} \sigma_{capt}v n\Delta t
\label{eq:fraction}
\end{equation}
where
\begin{equation}
\sigma_{capt} = 2\pi \biggl(\frac{85\pi}{6\sqrt{2}}\biggr)^{2/7}
G^2(M_1+M_2)^{10/7}(M_1M_2)^{2/7}v^{-18/7}
\label{eq:cross}
\end{equation}
is the capture cross section~\cite{Mouri:2002mc}. 
Here $G$ is the gravitational constant, $M_1$, $M_2$ are the two
BH masses, $v$ their relative velocity, $n$ the number density of the BHs,
and the speed of light is chosen
to be unity.
Such direct mergers had been considered 
in ~\cite{Bird:2016dcv,Clesse:2016ajp} though not within 
clusters formed due to Poissonian fluctuations. They were later on 
deemed subdominant due to the initial binary population.
The direct capture rate today may be estimated via
\begin{equation}
{\cal M} = \biggl(\frac{f_{direct}}{\Delta t}\biggr)
\biggl(\frac{f_M(M_{pbh})\rho^{avg}_{pbh}}{M_{pbh}}\biggr)
\label{eq:rate}
\end{equation} 
where $\rho_{pbh}^{avg}$ is the present cosmic average PBH mass density.
Using the above equations on the merger rate of two light PBHs we find
\begin{equation}
{\cal M}^d_{M_s-M_s}\approx 639\,\frac{1}{\rm Gpc^{3} yr} 
\biggl(\frac{M_s}{1 M_{\odot}}\biggr)^{-11/21}
\biggl(\frac{N_{cl}}{1300}\biggr)^{-137/84}
\label{eq:1_1}
\end{equation}
well below the current observational limit~\cite{Authors:2019qbw}.

\begin{figure}[htbp]
\centering
\includegraphics[width=0.48\textwidth]{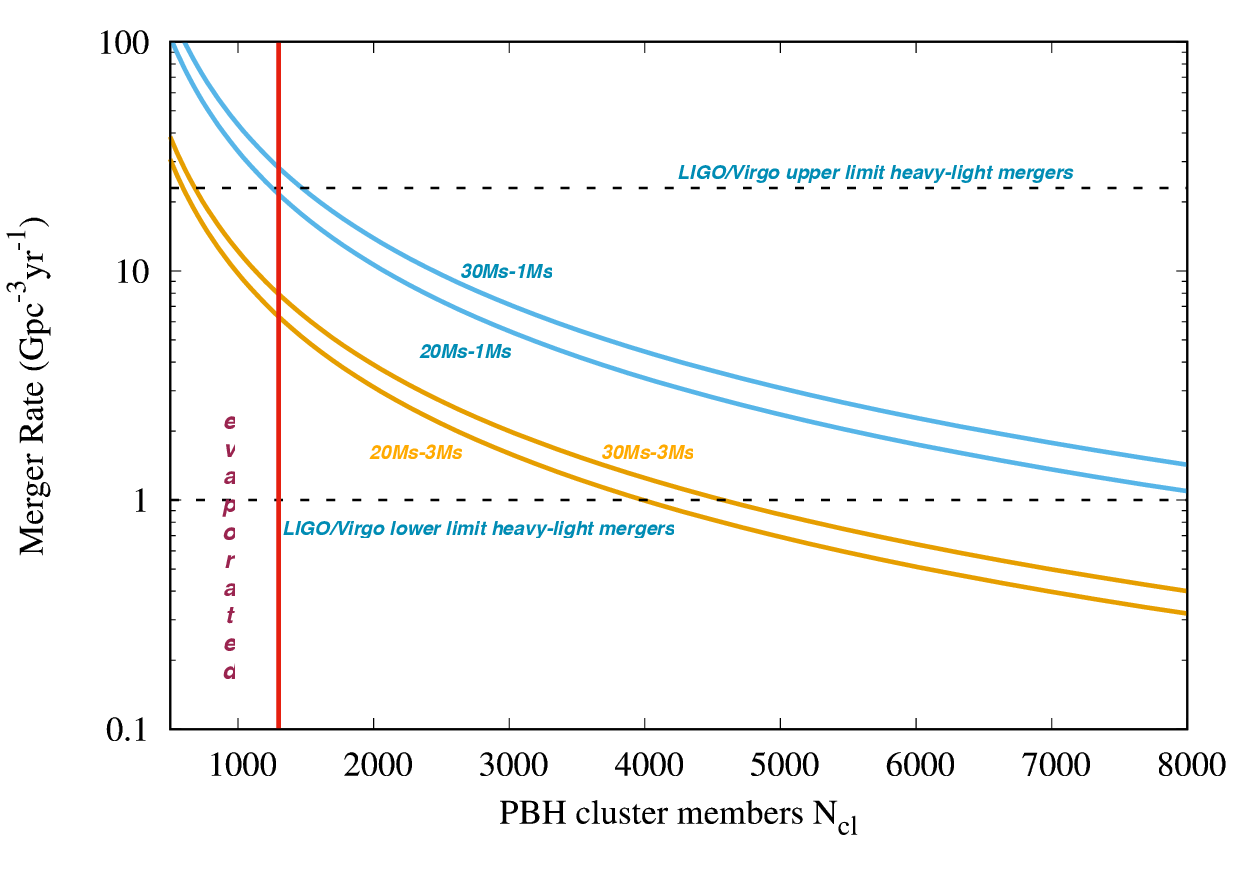}
\caption{\label{fig:merge_large_small} 
Merger rates of light PBHs with heavy PBHs as predicted
by Eq.~(\ref{eq:30_1_EOS}). Results for unpeaked 
inflationary perturbations are shown.
The two dotted lines present the by LIGO/Virgo
inferred upper and lower limit on this type of event, whereas the red 
line shows the critical $N_{cl}^{cr}$ of clusters which have evaporated
today. Four binary mass combinations are considered: 
$(M_1/M_{\odot},M_2/M_{\odot}) = (30,1), (30, 3), (20,1)\, {\rm and}\, (20,3)$, where the blue lines are for $M_2 = 1\,M_{\odot}$ and the orange lines are for $M_2 = 3\,M_{\odot}$ PBHs.}
\end{figure}

After formation clusters attain virial equilibrium quite rapidly.
In virial equilibrium, the typical
kinetic energies $E_k$ of heavy and light PBHs are the same, such that
$v (M_b) \equiv v_{M_b}\simeq v_{cl}\sqrt{M_s/M_b}$ where $v_{cl} =
v(M_s)$.
This leads to the sinking of heavy PBHs to the center of the cluster.
Adopting a simple top hat (i.e. constant density) model for the 
PBH clusters, the gravitational potential at radii $r < r_{cl}$
may be approximated by $V_p\simeq -(G M_{cl}M_b/r_{cl}) (r/r_{cl})^2$,
where $M_{cl}$ and $r_{cl}$ are total cluster mass and cluster virial 
radius, respectively.
Note that strictly speaking we would need to include self-consistently
the gravitational potential at the center of the cluster
provided by the heavy PBHs.
However since $f_M(M_b)$ is assumed to be small this effect is only of order 
unity and will here be omitted. 
Assuming virial equilibrium for the heavy $M_b$ PBHs as well, 
i.e. $E_k\simeq -V_p/2$, we may compute
the virial radius for the heavy PBHs to be 
$r^{vir}_{M_b}\simeq r_{cl}(v_{M_b}/v_{M_s}) \simeq r_{cl} \sqrt{M_s/M_b}$.
That implies, their number density is increased by a factor of
$\sqrt{M_b/M_s}^3$. Given this, one may compute the ratio of heavy to light
PBH number densities at the center of the cluster to be $f_M(M_b)\sqrt{M_b/M_s}$ which is 
much larger then the 
cosmic average of $f_M(M_b) (M_s/M_b)$. This estimate of the number density 
of the heavy PBHs, together with the direct merger cross section,
may be utilized to compute the present day merger rate of two large
PBHs with mass $M_b$. In particular, one may employ Eqs.(\ref{eq:fraction}),
(\ref{eq:cross}) and (\ref{eq:rate}) with the values of
$v = v_{Mb} \simeq v_{cl}\sqrt{M_s/M_b}$ and 
$n = n_{cl} f_M(M_b) \sqrt{M_b/M_s}^3$. This yields 
\begin{eqnarray}
{\cal M}^d_{M_b-M_b}\approx 152\,\frac{1}{\rm Gpc^{3} yr}
\biggl(\frac{f_M(M_b)}{10^{-2}}\biggr)^2 
\biggl(\frac{M_b}{30 M_{\odot}}\biggr)^{16/7} 
\label{eq:30_30_EOS}
\\
\nonumber
\times \biggl(\frac{M_s}{1 M_{\odot}}\biggr)^{-59/21}
\biggl(\frac{N_{cl}}{1300}\biggr)^{-137/84}
\end{eqnarray}
coming very close to that observed. However, it is important to note that
a typical cluster of $N_{cl} = 1300$ on average does not even 
include a single heavy PBH, due to the rarity of massive PBHs. Requiring that
the average number of massive PBHs in a cluster is two, i.e.
$(M_s/M_b)f_M(M_b)N_{cl}=2$, a cluster has to
be of size $N_{cl}=6000$ for an example $M_b = 30\,M_{\odot}$
implying a rate of 
${\cal M}^d_{30M_{\odot}-30M_{\odot}} \sim 12\,{\rm Gpc^{-3}yr^{-1}}$.
These values bracket very well those inferred from the LIGO/Virgo collaboration.

\begin{figure}[htbp]
\centering
\includegraphics[width=0.48\textwidth]{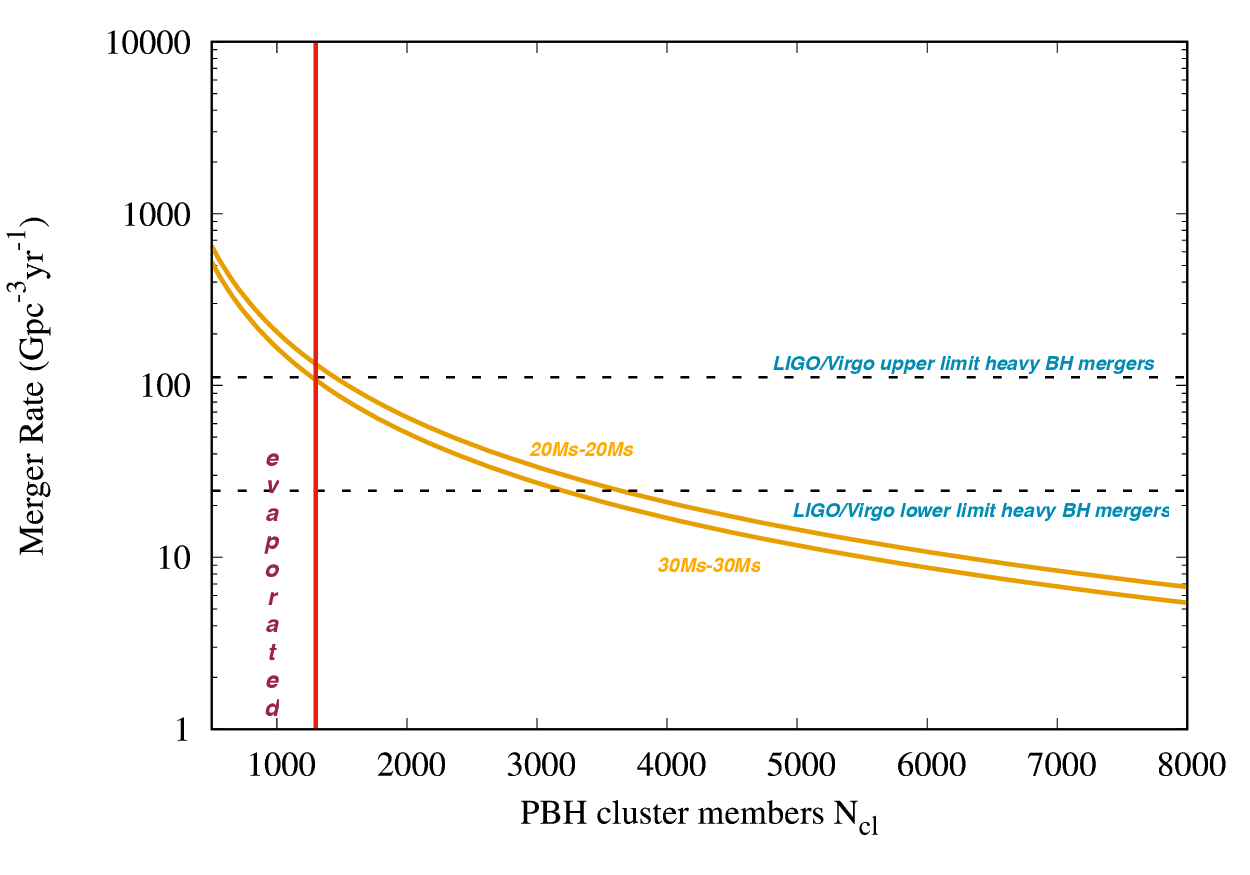}
\caption{\label{fig:destruction} Merger rates for inflationary 
perturbations with a peak on the QCD scale as predicted by
Eq.~(\ref{eq:30_30}). Shown is the equal mass
merger rate for heavy PBHs in the range
$20\, M_{\odot}\, <\, M_{pbh}\, <\, 30\, M_{\odot}$, as 
in Fig. (1).}
\end{figure} 

Next we calculate the merger rate of light PBHs with heavy PBHs. Here the
appropriate values to enter in Eq.~(\ref{eq:fraction})
are $v = v_{cl}$ since $v_{cl}\gg v_{Mb}$ and $n = n_{cl}$. This yields
\begin{eqnarray}
{\cal M}^d_{M_b-M_s}\approx 28.2\,\frac{1}{\rm Gpc^{3} yr}
\biggl(\frac{f_M(M_b)}{10^{-2}}\biggr) 
\biggl(\frac{M_b}{30 M_{\odot}}\biggr)^{5/7}
\label{eq:30_1_EOS} \\
\nonumber
\times \biggl(\frac{M_s}{1 M_{\odot}}\biggr)^{-26/21}
\biggl(\frac{N_{cl}}{1300}\biggr)^{-137/84}
\end{eqnarray}
Again requiring that on average there is at least one heavy 
PBH in the cluster, we find, for $M_s = 1\,M_{\odot}$ and
$M_b = 30M_{\odot}$ as an example
$N_{cl} = 3000$, yielding a rate ${\cal M}^d_{30M_{\odot}-1M_{\odot}} \sim 7.1\,{\rm Gpc^{-3}yr^{-1}}$. This should be compared to the by
LIGO/Virgo estimated rate ${\cal M}^d_{30M_{\odot}-1M_{\odot}} 
\sim 1 - 23\,{\rm Gpc^{-3}yr^{-1}}$ for GW190814 like events.
It is tantalizing that both merger rates in this simple estimate come so
close to the observed values. (see ~\cite{Vattis:2020iuz} for the merger 
rate if only one object is a PBH).

\vskip 0.15in
{\it PBH formation during the QCD epoch with an inflationary peak:}
\vskip 0.15in

We now consider scenarios where the pre-existing inflationary 
fluctuations necessary for PBH formation accidentally have a peak
on the mass scale $M_{pbh} = 30\, M_{\odot}$. 
We note, that in contrast to the prior scenario,
where their only was one free parameter, the fraction of dark matter in PBHs,
which we assumed to be unity, the scenario here has a second free 
parameter $M_{pbh}$. Simulations in ~\cite{Jedamzik:2020ypm} 
had only been performed for 
$M_{pbh} = 1\, M_{\odot}$, so, a priori we cannot be sure if similar
destruction rates of the initial population of very hard and eccentric
binaries apply. However, it is not too difficult to show, that physics
of binary destruction in clusters formed due to Poisson noise in the number
of PBHs, is independent of the mass scale $M_{pbh}$ and thus scale-invariant.
This applies also to 
the formation and evaporation of clusters. We may therefore
directly apply our findings 
in ~\cite{Jedamzik:2020ypm} to the current problem, implying that
current day mergers in PBH clusters due to the initial population are negligible. We may then use the above equations with $v = v_{cl}$ and
$n = n_{cl}$ to find the following rate 
\begin{equation}
{\cal M}^d_{M_b-M_b}\approx 108\,\frac{1}{\rm Gpc^{3} yr} 
\biggl(\frac{M_b}{30 M_{\odot}}\biggr)^{-11/21}
\biggl(\frac{N_{cl}}{1300}\biggr)^{-137/84}
\label{eq:30_30}
\end{equation}
Again, it is tantalizing that this rate is so close to that inferred 
by LIGO/Virgo. An estimate of the mergers of heavy $\sim 20-30\,M_{\odot}$
with light $\sim 1-3 M_{\odot}$ PBHs is difficult as there is no 
clear prediction, unlike in the QCD EOS
scenario. However, we may speculate
that there exists a broad and symmetric peak, and that there is some enhancement of PBH formation on 
the $\sim M_{\odot}$ PBH scale due to the QCD equation of
state. Since most PBHs are heavy they will dominate the properties
of clusters. In the attempt of light PBHs to attain kinetic energy
equilibrium in PBH clusters they will quickly reach velocities much beyond the virial velocity and evaporate. One may thus imagine current
galaxies and cluster of galaxies halos 
with a smooth light PBH background, but with
heavy PBH clusters, which dominate the gravitational potential of the
galaxy/cluster. We have estimated the light-heavy PBH
merger rate in this context. It would be given by the probability that
a light PBH enters into a PBH cluster and subsequently coalesces by a 
direct merger. We found the resultant merger rates orders of 
magnitude below those observed. Of course, it is conceivable that light
and heavy PBHs contribute almost similarly to the dark matter, going 
again a step towards the EOS scenario. In this 
case every heavy PBH cluster would be the center of a 
light PBH cluster with the latter having a much larger virial radius.
Thus merger rates could be parametrically enhanced.
We leave such considerations for future study.
\vskip 0.15 in
{\it The typical environment PBHs exist in at the current epoch:}
\vskip 0.15in

All these merger rates are displayed in Figs. (1), (2), and (3)
as a function of the number of cluster members $N_{cl}$ they reside in. 
It is important to realize that though $N_{cl}$ was chosen for the x-axis
in those figures, $N_{cl}$ is highly constrained. PBH clusters below
a critical $N_{cl}^{cr}$ have evaporated by now. In the spirit of 
hierarchical structure formation, clusters of smaller $N_{cl}^s$, but
with $N^s_{cl} > N_{cl}^{cr}$ are embedded in clusters of even 
larger $N^b_{cl}$. As a somewhat oversimplified example, one would
expect 10 clusters with $N^s_{cl} = 1300$ make up one cluster of
$N^b_{cl} = 13000$. The smallest clusters not yet evaporated
are therefore the basic building blocks. From detailed simulations of
CDM structure formation one has the experience that tidal forces are 
important in destroying some substructure. However, such arguments do
not play an important role for PBH clusters on small scales
formed from Poissonian noise, as the power spectrum on small scales is 
very different than that of CDM. PBH cluster densities scale as $n_{cl}\sim N_{cl}^{-3/2}$ such that the smallest clusters are much more dense and 
may not 
tidally be destroyed. Note that the existence of a population of high 
redshift clusters may explain a long-standing problem in cosmology, 
a significant excess in the cosmic infrared background on 
small scales~\cite{Kashlinsky:2006xg,Kashlinsky:2016sdv,Kashlinsky:2018mnu}.

\vskip 0.15in
{\it Mergers from PBHs which never entered clusters:}
\vskip 0.15in

The merger rate of the initial population of hard and eccentric
binaries is large~\cite{Nakamura:1997sm,Sasaki:2016jop}
\begin{equation}
{\cal M}^{ib}_{}\approx 1.25\times 10^6\,f_{\rm free}\,
f_M(M_{pbh})\,\frac{1}{\rm Gpc^{3} yr} 
\biggl(\frac{M_{pbh}}{1 M_{\odot}}\biggr)^{-32/37}
\end{equation}
implying that the fraction of PBHs which never entered a
cluster has to be smaller than 
\begin{eqnarray}
f_{\rm free}\,f_M(M_{pbh})\, & {}_{\sim}^< & \, 4.2\times 10^{-3}\quad {\rm for}\quad 
M_{pbh} = 1M_{\odot} \\ 
f_{\rm free}\,f_M(M_{pbh})\, & {}_{\sim}^< & \, 1.7\times 10^{-3}\quad  {\rm for}\quad
M_{pbh} = 30\,M_{\odot} 
\nonumber
\end{eqnarray}
in order to not exceed their respective
limits (see ~\cite{Vaskonen:2019jpv} for a related discussion). 
For PBH masses somewhat smaller than a solar mass these limits are
significantly relaxed due to a sensitivity loss of LIGO/Virgo. 
Such
small $f_{\rm free}$ (assuming $f_M(M_{pbh})\approx 1$)
are close to what one finds in CDM 
simulations~\cite{Stucker:2017nmi}. However,
values for $f_{\rm free}$ could be even smaller in PBH dark matter scenarios due
to the much earlier onset of structure formation and the initial zero
velocities of PBHs.
Nevertheless, it is important that $f_{\rm free}$ is determined by dedicated PBH structure formation simulations such as the one in ~\cite{Inman:2019wvr} 
\vskip 0.15in

Taken face value the QCD EOS scenario for PBH dark matter seems 
observationally challenged due to micro-lensing constraints in our galactic 
halo~\cite{Alcock:2000ph,Tisserand:2006zx} and due 
to high magnification micro-lensing observations in giant arcs due to caustic crossings~\cite{Oguri:2017ock}. 
On the other hand ~\cite{Alcock:2000ph}
reported a detection of a significant population of 
$\sim 0.5\, M_{\odot}$ unidentified objects, probably in the galactic halo.
A scenario of PBHs on the $\sim 30\, M_{\odot}$ scale due to an inflationary 
peak is less constrained by galactic micro-lensing, but still constrained by
observations of giant arcs.
As none if these constraints are overly stringent, and since such analysis
has to be adapted to the particularity of PBHs being in clusters, we believe
micro-lensing constraints deserve further scrutiny.

In concluding we would like to pause for a moment and have a look
from a distance at our findings. 
Scenarios where PBHs form during the QCD epoch have essentially
only one free parameter $f_{pbh}$, the contribution of such 
PBHs to the cosmic dark matter. Once this parameter has been fixed,
where we chose the natural Occam's razor 
value of $f_{pbh} = 1$, there remains no further
parametric freedom. Everything else is simply dictated by known physics.
In this highly constrained setting, PBHs formed during the QCD epoch
can (pre-) post-dict, the mass scale of $\sim 30\, M_{\odot}$ for PBHs
observed by LIGO/Virgo, the current merger rate of such PBHs
observed by LIGO/Virgo, the current merger rate of light PBHs 
with heavy ones observed by LIGO/Virgo, and the 
current non-observation of mergers on the fundamental 
$\sim 1\, M_{\odot}$ PBH scale. It may be that nature has not chosen this
pathway, but if not, has confronted us with an astonishing coincidence.
\vskip 0.15in

{\it Acknowledgements:}
I acknowledge useful exchanges with Yacine Ali-Hamoud, 
Krzysztof Belczynski, Derek Inman, Alexander Kashlinsky, 
Levon Pogosian, Ville Vaskonen,
and Hardi Veermae.


\begin{thebibliography}{50}%
\makeatletter
\providecommand \@ifxundefined [1]{%
 \@ifx{#1\undefined}
}%
\providecommand \@ifnum [1]{%
 \ifnum #1\expandafter \@firstoftwo
 \else \expandafter \@secondoftwo
 \fi
}%
\providecommand \@ifx [1]{%
 \ifx #1\expandafter \@firstoftwo
 \else \expandafter \@secondoftwo
 \fi
}%
\providecommand \natexlab [1]{#1}%
\providecommand \enquote  [1]{``#1''}%
\providecommand \bibnamefont  [1]{#1}%
\providecommand \bibfnamefont [1]{#1}%
\providecommand \citenamefont [1]{#1}%
\providecommand \href@noop [0]{\@secondoftwo}%
\providecommand \href [0]{\begingroup \@sanitize@url \@href}%
\providecommand \@href[1]{\@@startlink{#1}\@@href}%
\providecommand \@@href[1]{\endgroup#1\@@endlink}%
\providecommand \@sanitize@url [0]{\catcode `\\12\catcode `\$12\catcode
  `\&12\catcode `\#12\catcode `\^12\catcode `\_12\catcode `\%12\relax}%
\providecommand \@@startlink[1]{}%
\providecommand \@@endlink[0]{}%
\providecommand \url  [0]{\begingroup\@sanitize@url \@url }%
\providecommand \@url [1]{\endgroup\@href {#1}{\urlprefix }}%
\providecommand \urlprefix  [0]{URL }%
\providecommand \Eprint [0]{\href }%
\providecommand \doibase [0]{http://dx.doi.org/}%
\providecommand \selectlanguage [0]{\@gobble}%
\providecommand \bibinfo  [0]{\@secondoftwo}%
\providecommand \bibfield  [0]{\@secondoftwo}%
\providecommand \translation [1]{[#1]}%
\providecommand \BibitemOpen [0]{}%
\providecommand \bibitemStop [0]{}%
\providecommand \bibitemNoStop [0]{.\EOS\space}%
\providecommand \EOS [0]{\spacefactor3000\relax}%
\providecommand \BibitemShut  [1]{\csname bibitem#1\endcsname}%
\let\auto@bib@innerbib\@empty
\bibitem [{\citenamefont {Abbott}\ \emph {et~al.}(2016)\citenamefont {Abbott}
  \emph {et~al.}}]{Abbott:2016blz}%
  \BibitemOpen
  \bibfield  {author} {\bibinfo {author} {\bibfnamefont {B.}~\bibnamefont
  {Abbott}} \emph {et~al.} (\bibinfo {collaboration} {LIGO Scientific,
  Virgo}),\ }\href {\doibase 10.1103/PhysRevLett.116.061102} {\bibfield
  {journal} {\bibinfo  {journal} {Phys. Rev. Lett.}\ }\textbf {\bibinfo
  {volume} {116}},\ \bibinfo {pages} {061102} (\bibinfo {year} {2016})},\
  \Eprint {http://arxiv.org/abs/1602.03837} {arXiv:1602.03837 [gr-qc]}
  \BibitemShut {NoStop}%
\bibitem [{\citenamefont {Abbott}\ \emph
  {et~al.}(2019{\natexlab{a}})\citenamefont {Abbott} \emph
  {et~al.}}]{LIGOScientific:2018mvr}%
  \BibitemOpen
  \bibfield  {author} {\bibinfo {author} {\bibfnamefont {B.}~\bibnamefont
  {Abbott}} \emph {et~al.} (\bibinfo {collaboration} {LIGO Scientific,
  Virgo}),\ }\href {\doibase 10.1103/PhysRevX.9.031040} {\bibfield  {journal}
  {\bibinfo  {journal} {Phys. Rev. X}\ }\textbf {\bibinfo {volume} {9}},\
  \bibinfo {pages} {031040} (\bibinfo {year} {2019}{\natexlab{a}})},\ \Eprint
  {http://arxiv.org/abs/1811.12907} {arXiv:1811.12907 [astro-ph.HE]}
  \BibitemShut {NoStop}%
\bibitem [{\citenamefont {Abbott}\ \emph
  {et~al.}(2019{\natexlab{b}})\citenamefont {Abbott} \emph
  {et~al.}}]{LIGOScientific:2018jsj}%
  \BibitemOpen
  \bibfield  {author} {\bibinfo {author} {\bibfnamefont {B.}~\bibnamefont
  {Abbott}} \emph {et~al.} (\bibinfo {collaboration} {LIGO Scientific,
  Virgo}),\ }\href {\doibase 10.3847/2041-8213/ab3800} {\bibfield  {journal}
  {\bibinfo  {journal} {Astrophys. J. Lett.}\ }\textbf {\bibinfo {volume}
  {882}},\ \bibinfo {pages} {L24} (\bibinfo {year} {2019}{\natexlab{b}})},\
  \Eprint {http://arxiv.org/abs/1811.12940} {arXiv:1811.12940 [astro-ph.HE]}
  \BibitemShut {NoStop}%
\bibitem [{\citenamefont {Abbott}\ \emph {et~al.}(2020)\citenamefont {Abbott}
  \emph {et~al.}}]{Abbott:2020khf}%
  \BibitemOpen
  \bibfield  {author} {\bibinfo {author} {\bibfnamefont {R.}~\bibnamefont
  {Abbott}} \emph {et~al.} (\bibinfo {collaboration} {LIGO Scientific,
  Virgo}),\ }\href {\doibase 10.3847/2041-8213/ab960f} {\bibfield  {journal}
  {\bibinfo  {journal} {Astrophys. J.}\ }\textbf {\bibinfo {volume} {896}},\
  \bibinfo {pages} {L44} (\bibinfo {year} {2020})},\ \Eprint
  {http://arxiv.org/abs/2006.12611} {arXiv:2006.12611 [astro-ph.HE]}
  \BibitemShut {NoStop}%
\bibitem [{\citenamefont {Belczynski}\ \emph
  {et~al.}(2010{\natexlab{a}})\citenamefont {Belczynski}, \citenamefont
  {Bulik}, \citenamefont {Fryer}, \citenamefont {Ruiter}, \citenamefont
  {Vink},\ and\ \citenamefont {Hurley}}]{Belczynski:2009xy}%
  \BibitemOpen
  \bibfield  {author} {\bibinfo {author} {\bibfnamefont {K.}~\bibnamefont
  {Belczynski}}, \bibinfo {author} {\bibfnamefont {T.}~\bibnamefont {Bulik}},
  \bibinfo {author} {\bibfnamefont {C.~L.}\ \bibnamefont {Fryer}}, \bibinfo
  {author} {\bibfnamefont {A.}~\bibnamefont {Ruiter}}, \bibinfo {author}
  {\bibfnamefont {J.~S.}\ \bibnamefont {Vink}}, \ and\ \bibinfo {author}
  {\bibfnamefont {J.~R.}\ \bibnamefont {Hurley}},\ }\href {\doibase
  10.1088/0004-637X/714/2/1217} {\bibfield  {journal} {\bibinfo  {journal}
  {Astrophys. J.}\ }\textbf {\bibinfo {volume} {714}},\ \bibinfo {pages} {1217}
  (\bibinfo {year} {2010}{\natexlab{a}})},\ \Eprint
  {http://arxiv.org/abs/0904.2784} {arXiv:0904.2784 [astro-ph.SR]} \BibitemShut
  {NoStop}%
\bibitem [{\citenamefont {Belczynski}\ \emph
  {et~al.}(2010{\natexlab{b}})\citenamefont {Belczynski}, \citenamefont
  {Dominik}, \citenamefont {Bulik}, \citenamefont {O'Shaughnessy},
  \citenamefont {Fryer},\ and\ \citenamefont {Holz}}]{Belczynski:2010tb}%
  \BibitemOpen
  \bibfield  {author} {\bibinfo {author} {\bibfnamefont {K.}~\bibnamefont
  {Belczynski}}, \bibinfo {author} {\bibfnamefont {M.}~\bibnamefont {Dominik}},
  \bibinfo {author} {\bibfnamefont {T.}~\bibnamefont {Bulik}}, \bibinfo
  {author} {\bibfnamefont {R.}~\bibnamefont {O'Shaughnessy}}, \bibinfo {author}
  {\bibfnamefont {C.}~\bibnamefont {Fryer}}, \ and\ \bibinfo {author}
  {\bibfnamefont {D.~E.}\ \bibnamefont {Holz}},\ }\href {\doibase
  10.1088/2041-8205/715/2/L138} {\bibfield  {journal} {\bibinfo  {journal}
  {Astrophys. J. Lett.}\ }\textbf {\bibinfo {volume} {715}},\ \bibinfo {pages}
  {L138} (\bibinfo {year} {2010}{\natexlab{b}})},\ \Eprint
  {http://arxiv.org/abs/1004.0386} {arXiv:1004.0386 [astro-ph.HE]} \BibitemShut
  {NoStop}%
\bibitem [{\citenamefont {Broadhurst}\ \emph {et~al.}(2020)\citenamefont
  {Broadhurst}, \citenamefont {Diego},\ and\ \citenamefont
  {Smoot}}]{Broadhurst:2020cvm}%
  \BibitemOpen
  \bibfield  {author} {\bibinfo {author} {\bibfnamefont {T.}~\bibnamefont
  {Broadhurst}}, \bibinfo {author} {\bibfnamefont {J.~M.}\ \bibnamefont
  {Diego}}, \ and\ \bibinfo {author} {\bibfnamefont {G.~F.}\ \bibnamefont
  {Smoot}},\ }\href@noop {} {\  (\bibinfo {year} {2020})},\ \Eprint
  {http://arxiv.org/abs/2006.13219} {arXiv:2006.13219 [astro-ph.CO]}
  \BibitemShut {NoStop}%
\bibitem [{\citenamefont {Safarzadeh}\ and\ \citenamefont
  {Loeb}(2020)}]{Safarzadeh:2020ntc}%
  \BibitemOpen
  \bibfield  {author} {\bibinfo {author} {\bibfnamefont {M.}~\bibnamefont
  {Safarzadeh}}\ and\ \bibinfo {author} {\bibfnamefont {A.}~\bibnamefont
  {Loeb}},\ }\href@noop {} {\  (\bibinfo {year} {2020})},\ \Eprint
  {http://arxiv.org/abs/2007.00847} {arXiv:2007.00847 [astro-ph.HE]}
  \BibitemShut {NoStop}%
\bibitem [{\citenamefont {Khlopov}(2010)}]{Khlopov:2008qy}%
  \BibitemOpen
  \bibfield  {author} {\bibinfo {author} {\bibfnamefont {M.~Y.}\ \bibnamefont
  {Khlopov}},\ }\href {\doibase 10.1088/1674-4527/10/6/001} {\bibfield
  {journal} {\bibinfo  {journal} {Res. Astron. Astrophys.}\ }\textbf {\bibinfo
  {volume} {10}},\ \bibinfo {pages} {495} (\bibinfo {year} {2010})},\ \Eprint
  {http://arxiv.org/abs/0801.0116} {arXiv:0801.0116 [astro-ph]} \BibitemShut
  {NoStop}%
\bibitem [{\citenamefont {Carr}\ \emph {et~al.}(2020)\citenamefont {Carr},
  \citenamefont {Kohri}, \citenamefont {Sendouda},\ and\ \citenamefont
  {Yokoyama}}]{Carr:2020gox}%
  \BibitemOpen
  \bibfield  {author} {\bibinfo {author} {\bibfnamefont {B.}~\bibnamefont
  {Carr}}, \bibinfo {author} {\bibfnamefont {K.}~\bibnamefont {Kohri}},
  \bibinfo {author} {\bibfnamefont {Y.}~\bibnamefont {Sendouda}}, \ and\
  \bibinfo {author} {\bibfnamefont {J.}~\bibnamefont {Yokoyama}},\ }\href@noop
  {} {\  (\bibinfo {year} {2020})},\ \Eprint {http://arxiv.org/abs/2002.12778}
  {arXiv:2002.12778 [astro-ph.CO]} \BibitemShut {NoStop}%
\bibitem [{\citenamefont {Jedamzik}(1997)}]{Jedamzik:1996mr}%
  \BibitemOpen
  \bibfield  {author} {\bibinfo {author} {\bibfnamefont {K.}~\bibnamefont
  {Jedamzik}},\ }\href {\doibase 10.1103/PhysRevD.55.R5871} {\bibfield
  {journal} {\bibinfo  {journal} {Phys. Rev. D}\ }\textbf {\bibinfo {volume}
  {55}},\ \bibinfo {pages} {5871} (\bibinfo {year} {1997})},\ \Eprint
  {http://arxiv.org/abs/astro-ph/9605152} {arXiv:astro-ph/9605152} \BibitemShut
  {NoStop}%
\bibitem [{\citenamefont {Jedamzik}(1998)}]{Jedamzik:1998hc}%
  \BibitemOpen
  \bibfield  {author} {\bibinfo {author} {\bibfnamefont {K.}~\bibnamefont
  {Jedamzik}},\ }\href {\doibase 10.1016/S0370-1573(98)00067-2} {\bibfield
  {journal} {\bibinfo  {journal} {Phys. Rept.}\ }\textbf {\bibinfo {volume}
  {307}},\ \bibinfo {pages} {155} (\bibinfo {year} {1998})},\ \Eprint
  {http://arxiv.org/abs/astro-ph/9805147} {arXiv:astro-ph/9805147} \BibitemShut
  {NoStop}%
\bibitem [{\citenamefont {Jedamzik}\ and\ \citenamefont
  {Niemeyer}(1999)}]{Jedamzik:1999am}%
  \BibitemOpen
  \bibfield  {author} {\bibinfo {author} {\bibfnamefont {K.}~\bibnamefont
  {Jedamzik}}\ and\ \bibinfo {author} {\bibfnamefont {J.~C.}\ \bibnamefont
  {Niemeyer}},\ }\href {\doibase 10.1103/PhysRevD.59.124014} {\bibfield
  {journal} {\bibinfo  {journal} {Phys. Rev. D}\ }\textbf {\bibinfo {volume}
  {59}},\ \bibinfo {pages} {124014} (\bibinfo {year} {1999})},\ \Eprint
  {http://arxiv.org/abs/astro-ph/9901293} {arXiv:astro-ph/9901293} \BibitemShut
  {NoStop}%
\bibitem [{\citenamefont {Byrnes}\ \emph {et~al.}(2018)\citenamefont {Byrnes},
  \citenamefont {Hindmarsh}, \citenamefont {Young},\ and\ \citenamefont
  {Hawkins}}]{Byrnes:2018clq}%
  \BibitemOpen
  \bibfield  {author} {\bibinfo {author} {\bibfnamefont {C.~T.}\ \bibnamefont
  {Byrnes}}, \bibinfo {author} {\bibfnamefont {M.}~\bibnamefont {Hindmarsh}},
  \bibinfo {author} {\bibfnamefont {S.}~\bibnamefont {Young}}, \ and\ \bibinfo
  {author} {\bibfnamefont {M.~R.~S.}\ \bibnamefont {Hawkins}},\ }\href
  {\doibase 10.1088/1475-7516/2018/08/041} {\bibfield  {journal} {\bibinfo
  {journal} {JCAP}\ }\textbf {\bibinfo {volume} {08}},\ \bibinfo {pages} {041}
  (\bibinfo {year} {2018})},\ \Eprint {http://arxiv.org/abs/1801.06138}
  {arXiv:1801.06138 [astro-ph.CO]} \BibitemShut {NoStop}%
\bibitem [{\citenamefont {Carr}\ \emph {et~al.}(2019)\citenamefont {Carr},
  \citenamefont {Clesse}, \citenamefont {Garcia-Bellido},\ and\ \citenamefont
  {Kuhnel}}]{Carr:2019kxo}%
  \BibitemOpen
  \bibfield  {author} {\bibinfo {author} {\bibfnamefont {B.}~\bibnamefont
  {Carr}}, \bibinfo {author} {\bibfnamefont {S.}~\bibnamefont {Clesse}},
  \bibinfo {author} {\bibfnamefont {J.}~\bibnamefont {Garcia-Bellido}}, \ and\
  \bibinfo {author} {\bibfnamefont {F.}~\bibnamefont {Kuhnel}},\ }\href@noop {}
  {\  (\bibinfo {year} {2019})},\ \Eprint {http://arxiv.org/abs/1906.08217}
  {arXiv:1906.08217 [astro-ph.CO]} \BibitemShut {NoStop}%
\bibitem [{\citenamefont {Sobrinho}\ and\ \citenamefont
  {Augusto}(2020)}]{Sobrinho:2020cco}%
  \BibitemOpen
  \bibfield  {author} {\bibinfo {author} {\bibfnamefont {J.}~\bibnamefont
  {Sobrinho}}\ and\ \bibinfo {author} {\bibfnamefont {P.}~\bibnamefont
  {Augusto}},\ }\href@noop {} {\  (\bibinfo {year} {2020})},\ \Eprint
  {http://arxiv.org/abs/2005.10037} {arXiv:2005.10037 [astro-ph.CO]}
  \BibitemShut {NoStop}%
\bibitem [{\citenamefont {Borsanyi}\ \emph {et~al.}(2016)\citenamefont
  {Borsanyi} \emph {et~al.}}]{Borsanyi:2016ksw}%
  \BibitemOpen
  \bibfield  {author} {\bibinfo {author} {\bibfnamefont {S.}~\bibnamefont
  {Borsanyi}} \emph {et~al.},\ }\href {\doibase 10.1038/nature20115} {\bibfield
   {journal} {\bibinfo  {journal} {Nature}\ }\textbf {\bibinfo {volume}
  {539}},\ \bibinfo {pages} {69} (\bibinfo {year} {2016})},\ \Eprint
  {http://arxiv.org/abs/1606.07494} {arXiv:1606.07494 [hep-lat]} \BibitemShut
  {NoStop}%
\bibitem [{\citenamefont {Bhattacharya}\ \emph {et~al.}(2014)\citenamefont
  {Bhattacharya} \emph {et~al.}}]{Bhattacharya:2014ara}%
  \BibitemOpen
  \bibfield  {author} {\bibinfo {author} {\bibfnamefont {T.}~\bibnamefont
  {Bhattacharya}} \emph {et~al.},\ }\href {\doibase
  10.1103/PhysRevLett.113.082001} {\bibfield  {journal} {\bibinfo  {journal}
  {Phys. Rev. Lett.}\ }\textbf {\bibinfo {volume} {113}},\ \bibinfo {pages}
  {082001} (\bibinfo {year} {2014})},\ \Eprint {http://arxiv.org/abs/1402.5175}
  {arXiv:1402.5175 [hep-lat]} \BibitemShut {NoStop}%
\bibitem [{\citenamefont {Niemeyer}\ and\ \citenamefont
  {Jedamzik}(1998)}]{Niemeyer:1997mt}%
  \BibitemOpen
  \bibfield  {author} {\bibinfo {author} {\bibfnamefont {J.~C.}\ \bibnamefont
  {Niemeyer}}\ and\ \bibinfo {author} {\bibfnamefont {K.}~\bibnamefont
  {Jedamzik}},\ }\href {\doibase 10.1103/PhysRevLett.80.5481} {\bibfield
  {journal} {\bibinfo  {journal} {Phys. Rev. Lett.}\ }\textbf {\bibinfo
  {volume} {80}},\ \bibinfo {pages} {5481} (\bibinfo {year} {1998})},\ \Eprint
  {http://arxiv.org/abs/astro-ph/9709072} {arXiv:astro-ph/9709072} \BibitemShut
  {NoStop}%
\bibitem [{\citenamefont {Niemeyer}\ and\ \citenamefont
  {Jedamzik}(1999)}]{Niemeyer:1999ak}%
  \BibitemOpen
  \bibfield  {author} {\bibinfo {author} {\bibfnamefont {J.~C.}\ \bibnamefont
  {Niemeyer}}\ and\ \bibinfo {author} {\bibfnamefont {K.}~\bibnamefont
  {Jedamzik}},\ }\href {\doibase 10.1103/PhysRevD.59.124013} {\bibfield
  {journal} {\bibinfo  {journal} {Phys. Rev. D}\ }\textbf {\bibinfo {volume}
  {59}},\ \bibinfo {pages} {124013} (\bibinfo {year} {1999})},\ \Eprint
  {http://arxiv.org/abs/astro-ph/9901292} {arXiv:astro-ph/9901292} \BibitemShut
  {NoStop}%
\bibitem [{\citenamefont {Musco}\ \emph {et~al.}(2005)\citenamefont {Musco},
  \citenamefont {Miller},\ and\ \citenamefont {Rezzolla}}]{Musco:2004ak}%
  \BibitemOpen
  \bibfield  {author} {\bibinfo {author} {\bibfnamefont {I.}~\bibnamefont
  {Musco}}, \bibinfo {author} {\bibfnamefont {J.~C.}\ \bibnamefont {Miller}}, \
  and\ \bibinfo {author} {\bibfnamefont {L.}~\bibnamefont {Rezzolla}},\ }\href
  {\doibase 10.1088/0264-9381/22/7/013} {\bibfield  {journal} {\bibinfo
  {journal} {Class. Quant. Grav.}\ }\textbf {\bibinfo {volume} {22}},\ \bibinfo
  {pages} {1405} (\bibinfo {year} {2005})},\ \Eprint
  {http://arxiv.org/abs/gr-qc/0412063} {arXiv:gr-qc/0412063} \BibitemShut
  {NoStop}%
\bibitem [{\citenamefont {Ivanov}\ \emph {et~al.}(1994)\citenamefont {Ivanov},
  \citenamefont {Naselsky},\ and\ \citenamefont {Novikov}}]{Ivanov:1994pa}%
  \BibitemOpen
  \bibfield  {author} {\bibinfo {author} {\bibfnamefont {P.}~\bibnamefont
  {Ivanov}}, \bibinfo {author} {\bibfnamefont {P.}~\bibnamefont {Naselsky}}, \
  and\ \bibinfo {author} {\bibfnamefont {I.}~\bibnamefont {Novikov}},\ }\href
  {\doibase 10.1103/PhysRevD.50.7173} {\bibfield  {journal} {\bibinfo
  {journal} {Phys. Rev. D}\ }\textbf {\bibinfo {volume} {50}},\ \bibinfo
  {pages} {7173} (\bibinfo {year} {1994})}\BibitemShut {NoStop}%
\bibitem [{\citenamefont {Bullock}\ and\ \citenamefont
  {Primack}(1997)}]{Bullock:1996at}%
  \BibitemOpen
  \bibfield  {author} {\bibinfo {author} {\bibfnamefont {J.~S.}\ \bibnamefont
  {Bullock}}\ and\ \bibinfo {author} {\bibfnamefont {J.~R.}\ \bibnamefont
  {Primack}},\ }\href {\doibase 10.1103/PhysRevD.55.7423} {\bibfield  {journal}
  {\bibinfo  {journal} {Phys. Rev. D}\ }\textbf {\bibinfo {volume} {55}},\
  \bibinfo {pages} {7423} (\bibinfo {year} {1997})},\ \Eprint
  {http://arxiv.org/abs/astro-ph/9611106} {arXiv:astro-ph/9611106} \BibitemShut
  {NoStop}%
\bibitem [{\citenamefont {Dolgov}\ and\ \citenamefont
  {Silk}(1993)}]{Dolgov:1992pu}%
  \BibitemOpen
  \bibfield  {author} {\bibinfo {author} {\bibfnamefont {A.}~\bibnamefont
  {Dolgov}}\ and\ \bibinfo {author} {\bibfnamefont {J.}~\bibnamefont {Silk}},\
  }\href {\doibase 10.1103/PhysRevD.47.4244} {\bibfield  {journal} {\bibinfo
  {journal} {Phys. Rev. D}\ }\textbf {\bibinfo {volume} {47}},\ \bibinfo
  {pages} {4244} (\bibinfo {year} {1993})}\BibitemShut {NoStop}%
\bibitem [{\citenamefont {Sasaki}\ \emph {et~al.}(2016)\citenamefont {Sasaki},
  \citenamefont {Suyama}, \citenamefont {Tanaka},\ and\ \citenamefont
  {Yokoyama}}]{Sasaki:2016jop}%
  \BibitemOpen
  \bibfield  {author} {\bibinfo {author} {\bibfnamefont {M.}~\bibnamefont
  {Sasaki}}, \bibinfo {author} {\bibfnamefont {T.}~\bibnamefont {Suyama}},
  \bibinfo {author} {\bibfnamefont {T.}~\bibnamefont {Tanaka}}, \ and\ \bibinfo
  {author} {\bibfnamefont {S.}~\bibnamefont {Yokoyama}},\ }\href {\doibase
  10.1103/PhysRevLett.117.061101} {\bibfield  {journal} {\bibinfo  {journal}
  {Phys. Rev. Lett.}\ }\textbf {\bibinfo {volume} {117}},\ \bibinfo {pages}
  {061101} (\bibinfo {year} {2016})},\ \bibinfo {note} {[Erratum:
  Phys.Rev.Lett. 121, 059901 (2018)]},\ \Eprint
  {http://arxiv.org/abs/1603.08338} {arXiv:1603.08338 [astro-ph.CO]}
  \BibitemShut {NoStop}%
\bibitem [{\citenamefont {Ali-Haïmoud}\ \emph {et~al.}(2017)\citenamefont
  {Ali-Haïmoud}, \citenamefont {Kovetz},\ and\ \citenamefont
  {Kamionkowski}}]{Ali-Haimoud:2017rtz}%
  \BibitemOpen
  \bibfield  {author} {\bibinfo {author} {\bibfnamefont {Y.}~\bibnamefont
  {Ali-Haïmoud}}, \bibinfo {author} {\bibfnamefont {E.~D.}\ \bibnamefont
  {Kovetz}}, \ and\ \bibinfo {author} {\bibfnamefont {M.}~\bibnamefont
  {Kamionkowski}},\ }\href {\doibase 10.1103/PhysRevD.96.123523} {\bibfield
  {journal} {\bibinfo  {journal} {Phys. Rev. D}\ }\textbf {\bibinfo {volume}
  {96}},\ \bibinfo {pages} {123523} (\bibinfo {year} {2017})},\ \Eprint
  {http://arxiv.org/abs/1709.06576} {arXiv:1709.06576 [astro-ph.CO]}
  \BibitemShut {NoStop}%
\bibitem [{\citenamefont {Ballesteros}\ \emph {et~al.}(2018)\citenamefont
  {Ballesteros}, \citenamefont {Serpico},\ and\ \citenamefont
  {Taoso}}]{Ballesteros:2018swv}%
  \BibitemOpen
  \bibfield  {author} {\bibinfo {author} {\bibfnamefont {G.}~\bibnamefont
  {Ballesteros}}, \bibinfo {author} {\bibfnamefont {P.~D.}\ \bibnamefont
  {Serpico}}, \ and\ \bibinfo {author} {\bibfnamefont {M.}~\bibnamefont
  {Taoso}},\ }\href {\doibase 10.1088/1475-7516/2018/10/043} {\bibfield
  {journal} {\bibinfo  {journal} {JCAP}\ }\textbf {\bibinfo {volume} {10}},\
  \bibinfo {pages} {043} (\bibinfo {year} {2018})},\ \Eprint
  {http://arxiv.org/abs/1807.02084} {arXiv:1807.02084 [astro-ph.CO]}
  \BibitemShut {NoStop}%
\bibitem [{\citenamefont {Bringmann}\ \emph {et~al.}(2019)\citenamefont
  {Bringmann}, \citenamefont {Depta}, \citenamefont {Domcke},\ and\
  \citenamefont {Schmidt-Hoberg}}]{Bringmann:2018mxj}%
  \BibitemOpen
  \bibfield  {author} {\bibinfo {author} {\bibfnamefont {T.}~\bibnamefont
  {Bringmann}}, \bibinfo {author} {\bibfnamefont {P.~F.}\ \bibnamefont
  {Depta}}, \bibinfo {author} {\bibfnamefont {V.}~\bibnamefont {Domcke}}, \
  and\ \bibinfo {author} {\bibfnamefont {K.}~\bibnamefont {Schmidt-Hoberg}},\
  }\href {\doibase 10.1103/PhysRevD.99.063532} {\bibfield  {journal} {\bibinfo
  {journal} {Phys. Rev. D}\ }\textbf {\bibinfo {volume} {99}},\ \bibinfo
  {pages} {063532} (\bibinfo {year} {2019})},\ \Eprint
  {http://arxiv.org/abs/1808.05910} {arXiv:1808.05910 [astro-ph.CO]}
  \BibitemShut {NoStop}%
\bibitem [{\citenamefont {Raidal}\ \emph {et~al.}(2019)\citenamefont {Raidal},
  \citenamefont {Spethmann}, \citenamefont {Vaskonen},\ and\ \citenamefont
  {Veermäe}}]{Raidal:2018bbj}%
  \BibitemOpen
  \bibfield  {author} {\bibinfo {author} {\bibfnamefont {M.}~\bibnamefont
  {Raidal}}, \bibinfo {author} {\bibfnamefont {C.}~\bibnamefont {Spethmann}},
  \bibinfo {author} {\bibfnamefont {V.}~\bibnamefont {Vaskonen}}, \ and\
  \bibinfo {author} {\bibfnamefont {H.}~\bibnamefont {Veermäe}},\ }\href
  {\doibase 10.1088/1475-7516/2019/02/018} {\bibfield  {journal} {\bibinfo
  {journal} {JCAP}\ }\textbf {\bibinfo {volume} {02}},\ \bibinfo {pages} {018}
  (\bibinfo {year} {2019})},\ \Eprint {http://arxiv.org/abs/1812.01930}
  {arXiv:1812.01930 [astro-ph.CO]} \BibitemShut {NoStop}%
\bibitem [{\citenamefont {Vaskonen}\ and\ \citenamefont
  {Veermäe}(2020)}]{Vaskonen:2019jpv}%
  \BibitemOpen
  \bibfield  {author} {\bibinfo {author} {\bibfnamefont {V.}~\bibnamefont
  {Vaskonen}}\ and\ \bibinfo {author} {\bibfnamefont {H.}~\bibnamefont
  {Veermäe}},\ }\href {\doibase 10.1103/PhysRevD.101.043015} {\bibfield
  {journal} {\bibinfo  {journal} {Phys. Rev. D}\ }\textbf {\bibinfo {volume}
  {101}},\ \bibinfo {pages} {043015} (\bibinfo {year} {2020})},\ \Eprint
  {http://arxiv.org/abs/1908.09752} {arXiv:1908.09752 [astro-ph.CO]}
  \BibitemShut {NoStop}%
\bibitem [{\citenamefont {De~Luca}\ \emph {et~al.}(2020)\citenamefont
  {De~Luca}, \citenamefont {Franciolini}, \citenamefont {Pani},\ and\
  \citenamefont {Riotto}}]{DeLuca:2020qqa}%
  \BibitemOpen
  \bibfield  {author} {\bibinfo {author} {\bibfnamefont {V.}~\bibnamefont
  {De~Luca}}, \bibinfo {author} {\bibfnamefont {G.}~\bibnamefont
  {Franciolini}}, \bibinfo {author} {\bibfnamefont {P.}~\bibnamefont {Pani}}, \
  and\ \bibinfo {author} {\bibfnamefont {A.}~\bibnamefont {Riotto}},\
  }\href@noop {} {\  (\bibinfo {year} {2020})},\ \Eprint
  {http://arxiv.org/abs/2005.05641} {arXiv:2005.05641 [astro-ph.CO]}
  \BibitemShut {NoStop}%
\bibitem [{\citenamefont {Jedamzik}(2020)}]{Jedamzik:2020ypm}%
  \BibitemOpen
  \bibfield  {author} {\bibinfo {author} {\bibfnamefont {K.}~\bibnamefont
  {Jedamzik}},\ }\href@noop {} {\  (\bibinfo {year} {2020})},\ \Eprint
  {http://arxiv.org/abs/2006.11172} {arXiv:2006.11172 [astro-ph.CO]}
  \BibitemShut {NoStop}%
\bibitem [{\citenamefont {Afshordi}\ \emph {et~al.}(2003)\citenamefont
  {Afshordi}, \citenamefont {McDonald},\ and\ \citenamefont
  {Spergel}}]{Afshordi:2003zb}%
  \BibitemOpen
  \bibfield  {author} {\bibinfo {author} {\bibfnamefont {N.}~\bibnamefont
  {Afshordi}}, \bibinfo {author} {\bibfnamefont {P.}~\bibnamefont {McDonald}},
  \ and\ \bibinfo {author} {\bibfnamefont {D.}~\bibnamefont {Spergel}},\ }\href
  {\doibase 10.1086/378763} {\bibfield  {journal} {\bibinfo  {journal}
  {Astrophys. J. Lett.}\ }\textbf {\bibinfo {volume} {594}},\ \bibinfo {pages}
  {L71} (\bibinfo {year} {2003})},\ \Eprint
  {http://arxiv.org/abs/astro-ph/0302035} {arXiv:astro-ph/0302035} \BibitemShut
  {NoStop}%
\bibitem [{\citenamefont {Chisholm}(2006)}]{Chisholm:2005vm}%
  \BibitemOpen
  \bibfield  {author} {\bibinfo {author} {\bibfnamefont {J.~R.}\ \bibnamefont
  {Chisholm}},\ }\href {\doibase 10.1103/PhysRevD.73.083504} {\bibfield
  {journal} {\bibinfo  {journal} {Phys. Rev. D}\ }\textbf {\bibinfo {volume}
  {73}},\ \bibinfo {pages} {083504} (\bibinfo {year} {2006})},\ \Eprint
  {http://arxiv.org/abs/astro-ph/0509141} {arXiv:astro-ph/0509141} \BibitemShut
  {NoStop}%
\bibitem [{\citenamefont {Inman}\ and\ \citenamefont
  {Ali-Haïmoud}(2019)}]{Inman:2019wvr}%
  \BibitemOpen
  \bibfield  {author} {\bibinfo {author} {\bibfnamefont {D.}~\bibnamefont
  {Inman}}\ and\ \bibinfo {author} {\bibfnamefont {Y.}~\bibnamefont
  {Ali-Haïmoud}},\ }\href {\doibase 10.1103/PhysRevD.100.083528} {\bibfield
  {journal} {\bibinfo  {journal} {Phys. Rev. D}\ }\textbf {\bibinfo {volume}
  {100}},\ \bibinfo {pages} {083528} (\bibinfo {year} {2019})},\ \Eprint
  {http://arxiv.org/abs/1907.08129} {arXiv:1907.08129 [astro-ph.CO]}
  \BibitemShut {NoStop}%
\bibitem [{\citenamefont {Trashorras}\ \emph {et~al.}(2020)\citenamefont
  {Trashorras}, \citenamefont {García-Bellido},\ and\ \citenamefont
  {Nesseris}}]{Trashorras:2020mwn}%
  \BibitemOpen
  \bibfield  {author} {\bibinfo {author} {\bibfnamefont {M.}~\bibnamefont
  {Trashorras}}, \bibinfo {author} {\bibfnamefont {J.}~\bibnamefont
  {García-Bellido}}, \ and\ \bibinfo {author} {\bibfnamefont {S.}~\bibnamefont
  {Nesseris}},\ }\href@noop {} {\  (\bibinfo {year} {2020})},\ \Eprint
  {http://arxiv.org/abs/2006.15018} {arXiv:2006.15018 [astro-ph.CO]}
  \BibitemShut {NoStop}%
\bibitem [{\citenamefont {Binney}\ and\ \citenamefont {Tremaine}(2008)}]{BT08}%
  \BibitemOpen
  \bibfield  {author} {\bibinfo {author} {\bibfnamefont {J.}~\bibnamefont
  {Binney}}\ and\ \bibinfo {author} {\bibfnamefont {S.}~\bibnamefont
  {Tremaine}},\ }\href@noop {} {\emph {\bibinfo {title} {Galactic Dynamics}}},\
  \bibinfo {edition} {2nd}\ ed.\ (\bibinfo  {publisher} {Princeton University
  Press},\ \bibinfo {year} {2008})\BibitemShut {NoStop}%
\bibitem [{\citenamefont {Mouri}\ and\ \citenamefont
  {Taniguchi}(2002)}]{Mouri:2002mc}%
  \BibitemOpen
  \bibfield  {author} {\bibinfo {author} {\bibfnamefont {H.}~\bibnamefont
  {Mouri}}\ and\ \bibinfo {author} {\bibfnamefont {Y.}~\bibnamefont
  {Taniguchi}},\ }\href {\doibase 10.1086/339472} {\bibfield  {journal}
  {\bibinfo  {journal} {Astrophys. J. Lett.}\ }\textbf {\bibinfo {volume}
  {566}},\ \bibinfo {pages} {L17} (\bibinfo {year} {2002})},\ \Eprint
  {http://arxiv.org/abs/astro-ph/0201102} {arXiv:astro-ph/0201102} \BibitemShut
  {NoStop}%
\bibitem [{\citenamefont {Bird}\ \emph {et~al.}(2016)\citenamefont {Bird},
  \citenamefont {Cholis}, \citenamefont {Muñoz}, \citenamefont {Ali-Haïmoud},
  \citenamefont {Kamionkowski}, \citenamefont {Kovetz}, \citenamefont
  {Raccanelli},\ and\ \citenamefont {Riess}}]{Bird:2016dcv}%
  \BibitemOpen
  \bibfield  {author} {\bibinfo {author} {\bibfnamefont {S.}~\bibnamefont
  {Bird}}, \bibinfo {author} {\bibfnamefont {I.}~\bibnamefont {Cholis}},
  \bibinfo {author} {\bibfnamefont {J.~B.}\ \bibnamefont {Muñoz}}, \bibinfo
  {author} {\bibfnamefont {Y.}~\bibnamefont {Ali-Haïmoud}}, \bibinfo {author}
  {\bibfnamefont {M.}~\bibnamefont {Kamionkowski}}, \bibinfo {author}
  {\bibfnamefont {E.~D.}\ \bibnamefont {Kovetz}}, \bibinfo {author}
  {\bibfnamefont {A.}~\bibnamefont {Raccanelli}}, \ and\ \bibinfo {author}
  {\bibfnamefont {A.~G.}\ \bibnamefont {Riess}},\ }\href {\doibase
  10.1103/PhysRevLett.116.201301} {\bibfield  {journal} {\bibinfo  {journal}
  {Phys. Rev. Lett.}\ }\textbf {\bibinfo {volume} {116}},\ \bibinfo {pages}
  {201301} (\bibinfo {year} {2016})},\ \Eprint
  {http://arxiv.org/abs/1603.00464} {arXiv:1603.00464 [astro-ph.CO]}
  \BibitemShut {NoStop}%
\bibitem [{\citenamefont {Clesse}\ and\ \citenamefont
  {García-Bellido}(2017)}]{Clesse:2016ajp}%
  \BibitemOpen
  \bibfield  {author} {\bibinfo {author} {\bibfnamefont {S.}~\bibnamefont
  {Clesse}}\ and\ \bibinfo {author} {\bibfnamefont {J.}~\bibnamefont
  {García-Bellido}},\ }\href {\doibase 10.1016/j.dark.2017.10.001} {\bibfield
  {journal} {\bibinfo  {journal} {Phys. Dark Univ.}\ }\textbf {\bibinfo
  {volume} {18}},\ \bibinfo {pages} {105} (\bibinfo {year} {2017})},\ \Eprint
  {http://arxiv.org/abs/1610.08479} {arXiv:1610.08479 [astro-ph.CO]}
  \BibitemShut {NoStop}%
\bibitem [{\citenamefont {Abbott}\ \emph
  {et~al.}(2019{\natexlab{c}})\citenamefont {Abbott} \emph
  {et~al.}}]{Authors:2019qbw}%
  \BibitemOpen
  \bibfield  {author} {\bibinfo {author} {\bibfnamefont {B.}~\bibnamefont
  {Abbott}} \emph {et~al.} (\bibinfo {collaboration} {LIGO Scientific,
  Virgo}),\ }\href {\doibase 10.1103/PhysRevLett.123.161102} {\bibfield
  {journal} {\bibinfo  {journal} {Phys. Rev. Lett.}\ }\textbf {\bibinfo
  {volume} {123}},\ \bibinfo {pages} {161102} (\bibinfo {year}
  {2019}{\natexlab{c}})},\ \Eprint {http://arxiv.org/abs/1904.08976}
  {arXiv:1904.08976 [astro-ph.CO]} \BibitemShut {NoStop}%
\bibitem [{\citenamefont {Vattis}\ \emph {et~al.}(2020)\citenamefont {Vattis},
  \citenamefont {Goldstein},\ and\ \citenamefont
  {Koushiappas}}]{Vattis:2020iuz}%
  \BibitemOpen
  \bibfield  {author} {\bibinfo {author} {\bibfnamefont {K.}~\bibnamefont
  {Vattis}}, \bibinfo {author} {\bibfnamefont {I.~S.}\ \bibnamefont
  {Goldstein}}, \ and\ \bibinfo {author} {\bibfnamefont {S.~M.}\ \bibnamefont
  {Koushiappas}},\ }\href@noop {} {\  (\bibinfo {year} {2020})},\ \Eprint
  {http://arxiv.org/abs/2006.15675} {arXiv:2006.15675 [astro-ph.HE]}
  \BibitemShut {NoStop}%
\bibitem [{\citenamefont {Kashlinsky}\ \emph {et~al.}(2007)\citenamefont
  {Kashlinsky}, \citenamefont {Arendt}, \citenamefont {Mather},\ and\
  \citenamefont {Moseley}}]{Kashlinsky:2006xg}%
  \BibitemOpen
  \bibfield  {author} {\bibinfo {author} {\bibfnamefont {A.}~\bibnamefont
  {Kashlinsky}}, \bibinfo {author} {\bibfnamefont {R.}~\bibnamefont {Arendt}},
  \bibinfo {author} {\bibfnamefont {J.~C.}\ \bibnamefont {Mather}}, \ and\
  \bibinfo {author} {\bibfnamefont {S.}~\bibnamefont {Moseley}},\ }\href
  {\doibase 10.1086/513272} {\bibfield  {journal} {\bibinfo  {journal}
  {Astrophys. J. Lett.}\ }\textbf {\bibinfo {volume} {654}},\ \bibinfo {pages}
  {L5} (\bibinfo {year} {2007})},\ \bibinfo {note} {[Erratum: Astrophys.J.Lett.
  657, L131 (2007), Erratum: Astrophys.J. 657, L131 (2007)]},\ \Eprint
  {http://arxiv.org/abs/astro-ph/0612445} {arXiv:astro-ph/0612445} \BibitemShut
  {NoStop}%
\bibitem [{\citenamefont {Kashlinsky}(2016)}]{Kashlinsky:2016sdv}%
  \BibitemOpen
  \bibfield  {author} {\bibinfo {author} {\bibfnamefont {A.}~\bibnamefont
  {Kashlinsky}},\ }\href {\doibase 10.3847/2041-8205/823/2/L25} {\bibfield
  {journal} {\bibinfo  {journal} {Astrophys. J. Lett.}\ }\textbf {\bibinfo
  {volume} {823}},\ \bibinfo {pages} {L25} (\bibinfo {year} {2016})},\ \Eprint
  {http://arxiv.org/abs/1605.04023} {arXiv:1605.04023 [astro-ph.CO]}
  \BibitemShut {NoStop}%
\bibitem [{\citenamefont {Kashlinsky}\ \emph {et~al.}(2018)\citenamefont
  {Kashlinsky}, \citenamefont {Arendt}, \citenamefont {Atrio-Barandela},
  \citenamefont {Cappelluti}, \citenamefont {Ferrara},\ and\ \citenamefont
  {Hasinger}}]{Kashlinsky:2018mnu}%
  \BibitemOpen
  \bibfield  {author} {\bibinfo {author} {\bibfnamefont {A.}~\bibnamefont
  {Kashlinsky}}, \bibinfo {author} {\bibfnamefont {R.}~\bibnamefont {Arendt}},
  \bibinfo {author} {\bibfnamefont {F.}~\bibnamefont {Atrio-Barandela}},
  \bibinfo {author} {\bibfnamefont {N.}~\bibnamefont {Cappelluti}}, \bibinfo
  {author} {\bibfnamefont {A.}~\bibnamefont {Ferrara}}, \ and\ \bibinfo
  {author} {\bibfnamefont {G.}~\bibnamefont {Hasinger}},\ }\href {\doibase
  10.1103/RevModPhys.90.025006} {\bibfield  {journal} {\bibinfo  {journal}
  {Rev. Mod. Phys.}\ }\textbf {\bibinfo {volume} {90}},\ \bibinfo {pages}
  {025006} (\bibinfo {year} {2018})},\ \Eprint
  {http://arxiv.org/abs/1802.07774} {arXiv:1802.07774 [astro-ph.CO]}
  \BibitemShut {NoStop}%
\bibitem [{\citenamefont {Nakamura}\ \emph {et~al.}(1997)\citenamefont
  {Nakamura}, \citenamefont {Sasaki}, \citenamefont {Tanaka},\ and\
  \citenamefont {Thorne}}]{Nakamura:1997sm}%
  \BibitemOpen
  \bibfield  {author} {\bibinfo {author} {\bibfnamefont {T.}~\bibnamefont
  {Nakamura}}, \bibinfo {author} {\bibfnamefont {M.}~\bibnamefont {Sasaki}},
  \bibinfo {author} {\bibfnamefont {T.}~\bibnamefont {Tanaka}}, \ and\ \bibinfo
  {author} {\bibfnamefont {K.~S.}\ \bibnamefont {Thorne}},\ }\href {\doibase
  10.1086/310886} {\bibfield  {journal} {\bibinfo  {journal} {Astrophys. J.
  Lett.}\ }\textbf {\bibinfo {volume} {487}},\ \bibinfo {pages} {L139}
  (\bibinfo {year} {1997})},\ \Eprint {http://arxiv.org/abs/astro-ph/9708060}
  {arXiv:astro-ph/9708060} \BibitemShut {NoStop}%
\bibitem [{\citenamefont {Stucker}\ \emph {et~al.}(2018)\citenamefont
  {Stucker}, \citenamefont {Busch},\ and\ \citenamefont
  {White}}]{Stucker:2017nmi}%
  \BibitemOpen
  \bibfield  {author} {\bibinfo {author} {\bibfnamefont {J.}~\bibnamefont
  {Stucker}}, \bibinfo {author} {\bibfnamefont {P.}~\bibnamefont {Busch}}, \
  and\ \bibinfo {author} {\bibfnamefont {S.~D.}\ \bibnamefont {White}},\ }\href
  {\doibase 10.1093/mnras/sty815} {\bibfield  {journal} {\bibinfo  {journal}
  {Mon. Not. Roy. Astron. Soc.}\ }\textbf {\bibinfo {volume} {477}},\ \bibinfo
  {pages} {3230} (\bibinfo {year} {2018})},\ \Eprint
  {http://arxiv.org/abs/1710.09881} {arXiv:1710.09881 [astro-ph.CO]}
  \BibitemShut {NoStop}%
\bibitem [{\citenamefont {Alcock}\ \emph {et~al.}(2000)\citenamefont {Alcock}
  \emph {et~al.}}]{Alcock:2000ph}%
  \BibitemOpen
  \bibfield  {author} {\bibinfo {author} {\bibfnamefont {C.}~\bibnamefont
  {Alcock}} \emph {et~al.} (\bibinfo {collaboration} {MACHO}),\ }\href
  {\doibase 10.1086/309512} {\bibfield  {journal} {\bibinfo  {journal}
  {Astrophys. J.}\ }\textbf {\bibinfo {volume} {542}},\ \bibinfo {pages} {281}
  (\bibinfo {year} {2000})},\ \Eprint {http://arxiv.org/abs/astro-ph/0001272}
  {arXiv:astro-ph/0001272} \BibitemShut {NoStop}%
\bibitem [{\citenamefont {Tisserand}\ \emph {et~al.}(2007)\citenamefont
  {Tisserand} \emph {et~al.}}]{Tisserand:2006zx}%
  \BibitemOpen
  \bibfield  {author} {\bibinfo {author} {\bibfnamefont {P.}~\bibnamefont
  {Tisserand}} \emph {et~al.} (\bibinfo {collaboration} {EROS-2}),\ }\href
  {\doibase 10.1051/0004-6361:20066017} {\bibfield  {journal} {\bibinfo
  {journal} {Astron. Astrophys.}\ }\textbf {\bibinfo {volume} {469}},\ \bibinfo
  {pages} {387} (\bibinfo {year} {2007})},\ \Eprint
  {http://arxiv.org/abs/astro-ph/0607207} {arXiv:astro-ph/0607207} \BibitemShut
  {NoStop}%
\bibitem [{\citenamefont {Oguri}\ \emph {et~al.}(2018)\citenamefont {Oguri},
  \citenamefont {Diego}, \citenamefont {Kaiser}, \citenamefont {Kelly},\ and\
  \citenamefont {Broadhurst}}]{Oguri:2017ock}%
  \BibitemOpen
  \bibfield  {author} {\bibinfo {author} {\bibfnamefont {M.}~\bibnamefont
  {Oguri}}, \bibinfo {author} {\bibfnamefont {J.~M.}\ \bibnamefont {Diego}},
  \bibinfo {author} {\bibfnamefont {N.}~\bibnamefont {Kaiser}}, \bibinfo
  {author} {\bibfnamefont {P.~L.}\ \bibnamefont {Kelly}}, \ and\ \bibinfo
  {author} {\bibfnamefont {T.}~\bibnamefont {Broadhurst}},\ }\href {\doibase
  10.1103/PhysRevD.97.023518} {\bibfield  {journal} {\bibinfo  {journal} {Phys.
  Rev. D}\ }\textbf {\bibinfo {volume} {97}},\ \bibinfo {pages} {023518}
  (\bibinfo {year} {2018})},\ \Eprint {http://arxiv.org/abs/1710.00148}
  {arXiv:1710.00148 [astro-ph.CO]} \BibitemShut {NoStop}%
\end{thebibliography}

%

\end{document}